\documentclass[12pt]{article}
\usepackage{times}

\usepackage{epsfig}
\usepackage{graphicx,amssymb,amsbsy}
\usepackage{epsfig}

\newcounter{lastnote}
\newenvironment{scilastnote}{%
\setcounter{lastnote}{\value{enumiv}}%
\addtocounter{lastnote}{+1}%
\begin{list}%
{\arabic{lastnote}.}
{\setlength{\leftmargin}{.22in}}
{\setlength{\labelsep}{.5em}}}
{\end{list}}

\title{Formation of a Matter-Wave Bright Soliton}

\author {L.\,Khaykovich$^{1}$, F.\,Schreck$^{1}$, G.\,Ferrari$^{1,2}$,
T.\,Bourdel$^{1}$\\ J.\,Cubizolles$^{1}$, L. D. Carr$^{1}$,
Y. Castin$^{1}$, and C.\,Salomon$^{1}$\\ \\
\normalsize{$^{1}$Laboratoire Kastler Brossel, Ecole Normale
Sup\'erieure,}\\ \normalsize{ 24 rue Lhomond, 75231 Paris CEDEX 05,
France}\\ \normalsize{$^{2}$LENS-INFM, Largo E. Fermi 2, Firenze 50125,
Italy}\\ \\ }

\date{}

\begin{document}

\baselineskip24pt

\maketitle

\begin{abstract}
We report the production of matter-wave solitons in an ultra-cold $^7$Li
gas. The effective interaction between atoms in a Bose-Einstein
condensate is tuned with a Feshbach resonance from repulsive to
attractive before release in a one-dimensional optical
waveguide. Propagation of the soliton without dispersion over a
macroscopic distance of 1.1 mm is observed. A simple theoretical model
explains the stability region of the soliton.  These matter-wave
solitons open fascinating possibilities for future applications in
coherent atom optics, atom interferometry and atom transport.
\end{abstract}

\newpage
Solitons are localized waves that travel over long distances with
neither attenuation nor change of shape as their dispersion is
compensated by nonlinear effects.  Soliton research has been conducted
in fields as diverse as particle physics, molecular biology, geology,
oceanography, astrophysics and nonlinear optics. Perhaps the most
prominent application of solitons is in high rate telecommunications
using optical fibers \cite{SolitonSpecialissue}.

We use a Bose-Einstein condensate (BEC) of a dilute atomic gas of
lithium atoms as a macroscopic matter-wave to form a soliton.
Nonlinearity is provided by binary atomic interactions leading to the
mean-field potential $U(\vec{r}\,)= gn(\vec{r}\,)= 4\pi \hbar^2 a
n(\vec{r}\,)/m$, where $a$ is the scattering length, $n(\vec{r}\,)$ the
spatial density and $m$ the atomic mass. For $a<0$ the effective
interaction is attractive and a trapped BEC is only stable for a number
of atoms less than a critical number above which collapse occurs
\cite{Burnett,Hulet97, Wieman01}. When the BEC is confined in only two
directions, matter-waves have dispersion in the free direction due to
their kinetic energy, $E_{\rm kin} \propto k^2$, where $k$ is the atomic
wave vector. The balance between this dispersion and the attractive
mean-field energy can lead to the formation of bright solitons as shown
theoretically \cite{Zakharov72,BECSoliton, CarrSoliton}. Up to now, only
dark solitons have been observed in BEC's with repulsive interactions
($a>0$) \cite{Lewenstein99, Denschlag00}. These solitons are
characterized by a notch in the BEC density profile with a phase step
across the soliton center. They propagate within the BEC with a velocity
below the speed of sound, but so far are found to decay before reaching
the edge of the condensate.

We report on the formation of a matter-wave bright soliton, a freely
propagating self-bound atomic gas. The soliton is produced from a $^7$Li
BEC in the internal atomic state $|F=1,m_F=1\rangle$. In this state a
Feshbach resonance allows us to continuously tune the scattering length
from a positive to negative value by means of an applied magnetic field
\cite{Stoof93,Williams}, a requirement for the production of a bright
soliton.

In our experimental setup \cite{Mewes00,Schreck00,Schreck01}, $4\times
10^8$ $^7$Li atoms are loaded from a magneto-optical trap into a
strongly confining Ioffe-Pritchard (IP) magnetic trap. Atoms are in the
$|F=2, m_F=2\rangle$ state for which the scattering length is $a=-1.4$
nm. Evaporative cooling lowers the temperature from 2 mK to 10 $\mu$K
after which $\sim 6 \times 10^5$ atoms remain. Atoms are then
transferred into a far detuned optical dipole trap at the intersection
of two Nd:YAG gaussian laser beams (Fig. 1) with common waists of 38(3)
$\mu$m \cite{Grimm}. The 9.5 W laser power is split between the two
beams using two acousto-optic modulators.

\begin{figure}[!hb]
\begin{center}
\includegraphics[width=8cm]{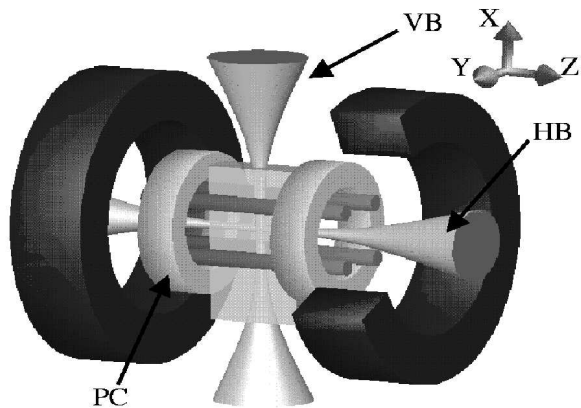}
\caption{Experimental setup for soliton production.  $^7$Li atoms are
evaporatively cooled in a Ioffe-Pritchard magnetic trap and transferred
into a crossed optical dipole trap in state $|F=1,m_F=1>$ where they
Bose condense. Magnetic tuning of the scattering length to positive,
zero and negative values is performed with the two pinch coils
(PC). Switching-off the vertical trapping beam (VB) allows propagation
of a soliton in the horizontal 1D waveguide (HB). Absorption images of
solitons and BEC's are recorded on a CCD camera in the $x,z$ plane.}
\end{center}
\end{figure}

The transfer from the magnetic trap to the optical trap is done in two
steps. First, the power of the YAG beams is ramped over 200 ms to a
value such that the radial oscillation frequency of the atoms is 1.8 kHz
in the vertical beam and 3.3 kHz in the horizontal beam which matches
that of the IP trap. Second, the magnetic trap is slowly turned off over
200 ms, keeping only a 5 G bias field.  The transfer efficiency is
nearly 100\%. Then, transfer from the state $|F=2,m_F=2\rangle$ to the
state $|F=1,m_F=1\rangle$ is performed by rapid adiabatic passage with a
microwave frequency sweep scanning 1 MHz in 10 ms around 820\,MHz. The
transfer efficiency is better than 95\%. Among all $^7$Li hyperfine
states which can be trapped in the dipole trap, $|F=1,m_F=1\rangle$ is
particularly useful as it is the lowest energy state in which 2-body
losses, which are relatively strong in the state $|F=2,m_F=2\rangle$
\cite{Schreck00}, are completely suppressed.  Furthermore, this state is
predicted to have a Feshbach resonance near $725$~G \cite{Williams},
allowing magnetic tuning of the scattering length (Fig. 2). An
adjustable magnetic field is produced by the pinch coils of our IP
trap. Their inductance is small so that their current can be changed on
a time scale shorter than $\sim 200\,\mu$s. As in previous work on
$^{23}$Na and $^{85}$Rb \cite{Ketterle98, Wieman00}, we locate the
$^7$Li Feshbach resonance through observation of a dramatic loss of
trapped atoms that we experimentally identify as due to 3-body
recombination. The resonance position is found at $720(15)$\,G, in good
agreement with theory (725 G) \cite{Williams}.

\begin{figure}[!hb]
\begin{center}
\includegraphics[width=8cm]{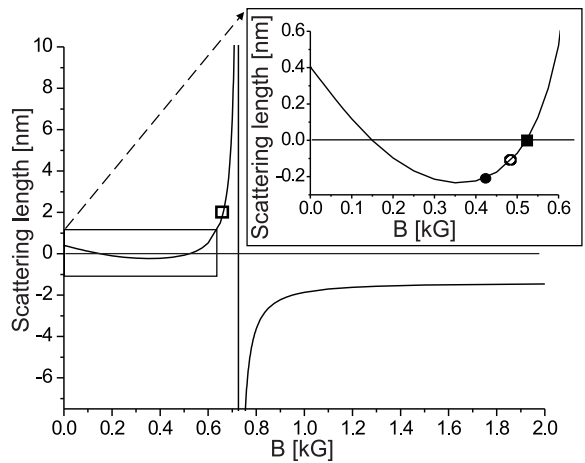}
\caption{ Predicted magnetic field dependence of the scattering length
$a$ for $^7$Li in state $|F=1,m_F =1\rangle$ \cite{Williams}. Insert:
expanded view of the $[0\,-0.6\,]$~kG interval with the various values
of $a$ used to study soliton formation.  Open square: initial
BEC. Square: Ideal BEC gas. Open circle: attractive gas. Dark circle:
soliton.}
\end{center}
\end{figure}

We then produce a $^7$Li BEC in the crossed dipole trap by forced
evaporation achieved by lowering the depth of the optical trapping
potential \cite{Chapman01}. Between $B=0$ and $B=590$\,G the scattering
length is small ($|a| \leq 0.4$~nm), hindering efficient evaporative
cooling (Fig. 2). Therefore, we operate at a magnetic field of $665$\,G
in the wing of the Feshbach resonance where $a \simeq +2.1$\,nm and
where 3-body losses remain moderate. The horizontal (vertical) optical
power is lowered from $5.5$~W ($1.5~$W) to $1.15$~W ($0.9$~W) in
$100$~ms, and then to $0.27$~W ($0.19$~W) in $150$ ms. A condensate with
$N\sim 2\times 10^4$ atoms, about half of the total number of atoms, is
obtained in a nearly isotropic trap where atoms have oscillation
frequencies of $710\, ,1000\, ,710$\,Hz along x, y, z. We then tune the
scattering length to zero to reduce 3-body losses.

In order to transform the BEC into a bright soliton, the trapping
geometry is adiabatically deformed to a cylindrical geometry obtained by
keeping only the horizontal trapping beam. To ensure adiabatic
deformation of the condensate, the vertical beam power is ramped down to
$3$~mW in 200\,ms, which reduces the axial oscillation frequency of the
atoms to $\omega_{z}\simeq 2\pi\times 50\,$Hz while the radial
oscillation frequency remains $\omega_{\bot}=2\pi\times 710\,$Hz. The
effective interaction is then tuned through changes in the magnetic
field in 50\,ms.  Finally, switching off the vertical beam with a
mechanical shutter releases the BEC into the horizontal 1D waveguide.
In the axial direction the coils that are used to provide the offset
field produce a slightly expulsive harmonic potential for the state
$|F=1,m_F=1\rangle$ which overcomes the dipole trap. The resulting axial
force on the atoms is conveniently written as $-m\omega_z^2 z$ where the
frequency $\omega_z$ is now imaginary. Typically $\omega_{z}=2i\pi\times
78\,$Hz for $B=520$\,G. After an adjustable evolution time in the
horizontal guide, the bias magnetic field is turned off and $400\,\mu$s
later an absorption image is recorded (Fig.~3) where the formation of
the soliton is seen.

\begin{figure}[!ht]

\begin{center}
\includegraphics[width=8cm]{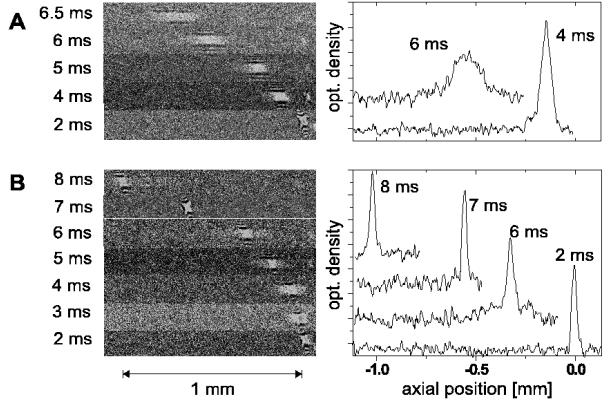}
\caption{ Absorption images at variable delays after switching off the
vertical trapping beam. Propagation of an ideal BEC gas (A) and of a
soliton (B) in the horizontal 1D waveguide in presence of an expulsive
potential. Propagation without dispersion over $1.1$~mm is a clear
signature of a soliton. Corresponding axial profiles integrated over the
vertical direction.}
\end{center}
\end{figure}

We compare the evolution of an ideal gas (Fig.~3A)), $a\simeq 0$
for B=520\,G, with a gas with attractive interactions
(Fig.~3B,SFig.~3B), $a=-0.21\,$nm for $B=425$\,G. In both cases
the cloud drifts towards the left because of a small offset,
$\simeq 50\,\mu$m\,, between the maximum of the expulsive
potential and the initial position of the atoms. The width of the
expanding cloud in the horizontal waveguide is considerably
broader in the non-interacting case, while for all times of
observation the soliton width remains equal to the resolution
limit of our imaging system, $9(1)\,\mu$m axially
\cite{resolution}. The cloud contains $6(2)\times 10^3$ atoms and
propagates over a distance of 1.1 mm without any detectable
dispersion, a clear signature of a bright soliton {\bf
\cite{curv}}. No decay of the soliton is observed in the $10\,$ms
it remains in the detection region. A substantial fraction of
atoms, $\simeq 2/3$, remains in a non-condensed pedestal around
the soliton, clearly visible for intermediate propagation times in
the guide.

We then made measurements of the wave-packet size versus propagation
time for three values of the scattering length: $a\simeq 0, a\simeq
-0.11$\,nm and $a\simeq -0.21$\,nm (Fig.~4). For $a\simeq 0$ (Fig.~4A),
the interaction between atoms is negligible and the size of the cloud is
governed by the expansion of the initial condensate distribution under
the influence of the negative curvature of the axial potential. The
measured size is in excellent agreement with the predicted size of a
non-interacting gas subjected to an expulsive harmonic potential: taking
the curvature as a fit parameter (solid line in Fig.~4A), we find
$\omega_{z} = 2i\pi\times78(3)$\, Hz, which agrees with the expected
value of the curvature produced by the pinch coils \cite{Schreck01}. For
$a=-0.11$ nm and $B=487\,$G the size of the wave-packet is consistently
below that of a non-interacting gas (Fig.~4B : solid line). Attractive
interactions reduce the size of the atomic cloud but are not strong
enough to stabilize the soliton against the expulsive potential. When
$a$ is further decreased to $-0.21$~nm the measured size of the wave
packet no longer changes as a function of guiding time, indicating
propagation without dispersion even in presence of the expulsive
potential (Fig.~4C).

\begin{figure}[!hb]
\begin{center}
\includegraphics[width=8cm]{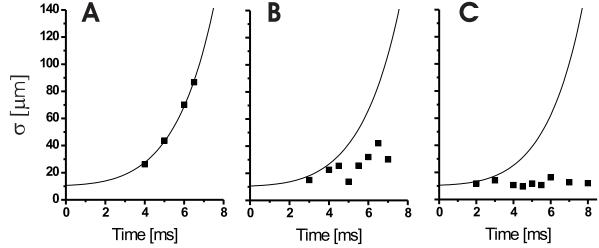}
\caption{Measured r.m.s size of the atomic wave-packet Gaussian fit as a
function of propagation time in the waveguide. (A): $a = 0$, ideal gas
case; (B): $a = -0.11\,$ nm; (C): $a = -0.21\,$nm; solid lines:
calculated expansion of a non-interacting gas in the expulsive
potential.}
\end{center}
\end{figure}

To theoretically analyze the stability of the soliton, we
introduce the three-dimensional Gross-Pitaevskii energy functional
\begin{eqnarray}
E_{\rm GP}   &=& \int {\mathrm d}^{3}r\frac{\hbar^{2}}{2\,m}{\left|\nabla\Psi\left(\vec{r}\right)\right|}^{2}+\frac{N g}{2}{\left|
\Psi\left(\vec{r}\right)\right|}^{4} \nonumber \\
             &+& \frac{1}{2}m\left[\omega^{2}_{\perp}\left(x^2+y^2\right)+\omega^{2}_{z}z^{2}\right]
{\left|\Psi\left(\vec{r}\right)\right|}^{2}, \label{eqn:gpe}
\end{eqnarray}
where the condensate wave-function $\Psi$ is normalized to one. In
Eq.~\ref{eqn:gpe} the first term is the kinetic energy responsible for
dispersion, the second term is the interaction energy, which in the
present case of attractive effective interactions ($g<0)$ causes the
wave-function to sharpen, and the third term is the external potential
energy. We introduce the following two-parameter variational ansatz to
estimate minimal energy states of $E_{\rm GP}$:
\begin{equation}
\Psi(\vec{r}\,)=\frac{1}{\sqrt{2\pi \sigma_{\bot}^2 l_z}}\,
\frac{1}{\cosh(z/l_z)}
\exp\left(-\frac{x^2+y^2}{2\sigma_{\bot}^2}\right)\,,
\label{eqn:sol}
\end{equation}
where $\sigma_{\bot}$ and $l_z$ are the radial and axial widths of the
wave-function. The functional form of the well-known 1D soliton has been
chosen for the longitudinal direction \cite{Zakharov72}, while in the
transverse direction a Gaussian ansatz is the optimal one for harmonic
confinement. For each $l_z$ we minimize the mean energy over
$\sigma_\bot$; the resulting function of $l_z$ is plotted (Fig.~5) for
various values of the parameter $Na/a_\bot^{\rm ho}$ where $ a_\bot^{\rm
ho}=(\hbar/m\omega_\bot)^{1/2}$. For very small axial sizes, the
interaction energy becomes on the order of $-\hbar\omega_\perp$ and the
gas loses its quasi-1D nature and collapses \cite{Hulet97,
Wieman01}. For very large axial sizes the expulsive potential energy
dominates and pulls the wave function apart. For intermediate sizes,
attractive interactions balance both the dispersion and the effect of
the expulsive potential; the energy presents a local minimum (solid line
in Fig.~5). This minimum supports a macroscopic quantum bound
state. However, it exists only within a narrow window of the parameter
$Na/a_\bot^{\rm ho}$. In our experiments $\omega_{\bot}=2\pi\times
710$~Hz and $\omega_z = 2i\pi\times 70$~Hz for $B=420$\,G, so that
$a_\bot^{\rm ho}=1.4\,\mu$m; for $N|a|$ larger than $(N|a|)_c =
1.105\,\mu$m, a collapse occurs (dashed curve in Fig.~5), while for
$N|a|$ smaller than $(N|a|)_e = 0.88\,\mu$m the expulsive potential
causes the gas to explode axially (dotted curve in Fig.~5).

\begin{figure}[!hb]
\begin{center}
\includegraphics[width=8cm]{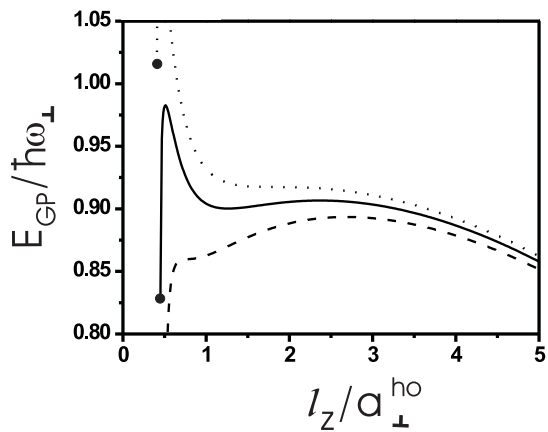}
\caption{Theoretical energy diagram of an attractive Bose gas subjected
to an expulsive potential for $\omega_z/\omega_{\perp}=i\times
70/710$. The energy as a function of the axial size after minimization
over the tranverse size is shown for three values of $N|a|$: within the
stability window (solid curve); at the critical point for explosion
$(N|a|)_e$ (dotted curve); and at the critical point for collapse
$(N|a|)_c$ (dashed curve). End points of the curves indicate collapse,
i.e. $\sigma_{\perp}=0$.}
\end{center}
\end{figure}

For our experimental conditions and $a=-0.21\,$nm, the number of atoms
which allows the soliton to be formed is $4.2\times 10^3 \leq N\leq
5.2\times 10^3$, in good agreement with our measured number $6(2)\times
10^3$. The expected axial size of the soliton is $l_z\simeq 1.7\,\mu$m,
which is below the current resolution limit of our imaging system. To
verify the presence of a critical value of $(Na)_e$ needed to stabilize
the soliton, we have performed the measurements with the same $a$ but
with a reduced number of atoms, $N=2\times 10^{3}$. At 8 ms guiding time
the axial size of the wave packet increased to 30 $\mu$m indicating that
no soliton was formed.

One may speculate as to the formation dynamics of the soliton in the
elongated trap before its release in the optical waveguide. As the atom
number in the initial BEC, $2\times 10^{4}$, is about three times larger
than the measured atom number in the soliton, it is likely that during
the 50 ms phase where $a$ is changed from 0 to negative values one or
several collapses occur until the critical number for a stable BEC is
reached. Indeed, the collapse time constant is predicted to be much less
than 50\, ms for our experimental conditions \cite{Shlyapnikov}. During
the transfer into the 1D waveguide the BEC is transformed into a soliton
and the non-condensed cloud is clearly observed at guiding times up to 6
ms as a broader background distribution. Non-adiabatic projection of the
BEC from the confining onto the expulsive potential is expected to play
a negligible role here according to numerical simulations
\cite{Carr}. At longer times the non-condensed atoms spread apart and
become undetectable. Thus during the propagation phase the soliton
decouples itself from the non-condensed fraction, resulting in a nearly
pure soliton.

Finally, removal of the expulsive axial potential will allow us to
significantly extend the stability domain towards lower values of $N|a|$
and longer observation times. The soliton size could then be measured in
situ, as well as its lifetime. The study of coherence properties of
solitons and of binary collisions between solitons are immediate
extensions of the present work.

\begin{scilastnote}
\item
We are grateful to K. Corwin, M. Olshanii, G. Shlyapnikov, C. Williams,
V. Venturi and B. Esry for important contributions to this work and to
J. Dalibard and C. Cohen-Tannoudji for useful discussions. F.\,S. was
supported by the DAAD, G.\,F. by the EU network CT 2000-00165 CAUAC, and
L.D.C. by the NSF MPS-DRF 0104447. This work was supported by CNRS,
Coll\`ege de France and R\'egion Ile de France.  Laboratoire Kastler
Brossel is {\it Unit\'e de recherche de l'Ecole Normale Sup\'erieure et
de l'Universit\'e Pierre et Marie Curie, associ\'ee au CNRS}.
\end{scilastnote}


\begin{thebibliography}{99}
\bibitem{curv} Over this distance, the change in magnetic field due to
the curvature is $0.1$~G, therefore the change in the scattering length
is negligible (Fig. 2).
\end{thebibliography}
\end{document}